\documentstyle[aps,epsfig,multicol]{revtex}

\begin{document}

\def\bea{\begin{eqnarray}}
\def\eea{\end{eqnarray}}
\def\a{\alpha}
\def\D{\langle l \rangle}
\def\p{\partial} 


\title{Nonequilibrium Phase Transition in a Model of Diffusion, 
Aggregation and Fragmentation}

\author{Satya N. Majumdar$^1$, Supriya Krishnamurthy$^2$, and Mustansir
Barma$^1$}
\address{$^1$ Department of Theoretical Physics, Tata
Institute of Fundamental Research\\
Homi Bhabha Road, Bombay 400005, India. \\
$^2$ Laboratoire Physique et M\'ecanique des Milieux
H\'et\'erogenes,\\
Ecole Sup\'erieure de Physique et Chimie Industrielles,\\
10 rue Vauquelin, 75231 Paris cedex 05, France.\\}
\maketitle
\widetext



\begin{abstract}
We study the nonequilibrium phase transition in a model of aggregation
of masses allowing for diffusion,
aggregation on contact and fragmentation. The model undergoes a dynamical
phase transition in all dimensions. The steady state mass distribution
decays exponentially for large mass in one phase. On the contrary,
in the other phase it has a power law tail and in addition 
an infinite aggregate. The model is solved exactly
within a mean field approximation which keeps track of the
distribution of masses. In one dimension, by mapping to an equivalent
lattice gas model, exact steady states are obtained in two extreme limits
of the parameter space.
Critical exponents and the phase diagram are obtained numerically in one
dimension. We also study the time dependent fluctuations in an equivalent
interface model in $(1+1)$ dimension and compute the roughness exponent
$\chi$ and the dynamical exponent $z$ analytically in some limits
and numerically otherwise. Two new fixed points of interface fluctuations
in $(1+1)$ dimension are identified. We also generalize our model to
include arbitrary fragmentation kernels and solve the steady states
exactly for some special choices of these kernels via mappings to
other solvable models of statistical mechanics.  

\end{abstract}



\begin{multicols}{2}
\section{Introduction}

By now, there is a fairly good understanding of 
the nature and properties of phase transitions in systems in thermal
equilibrium, as one changes the strengths of external fields 
such as temperature,
pressure or magnetic field. On the other hand there
is a wide variety of inherently {\em nonequilibrium} systems in nature
whose steady states are not described by the equilibrium Gibbs
distribution, but
are instead determined by the underlying microscopic
dynamical processes. The steady states of such systems may undergo
nonequilibrium phase transitions as one changes the rates of the
underlying dynamical processes. As compared to their equilibrium
counterparts, these nonequilibrium steady states and the transitions
between them are much less understood owing to the lack of a general
framework. It is therefore necessary to study simple
models of nonequilibrium processes, both in order to
discover new types of transitions as well as to understand the mechanisms 
which give rise to them.

In this paper, we study nonequilibrium phase transitions in an
important class of systems which involve the microscopic processes of
diffusion, aggregation upon contact and fragmentation of masses. These
processes arise in a variety of physical settings, for example, in the
formation of colloidal suspensions\cite{White}, polymer
gels\cite{Ziff}, river networks\cite{river}, aerosols and
clouds\cite{Fried}. They also enter in an important way in surface
growth phenomena involving island formation\cite{Lewis}. 
Below we introduce a simple lattice model incorporating these
microscopic processes and study the nonequilibrium steady states and
the transitions between them both analytically within mean field
theory and numerically in one dimension. Some of the results of this
paper have been reported earlier in a shorter version\cite{MKB1}

The paper is organized as follows. In section II, we define the model
and summarize our main results.  In section III, we solve the mean
field theory exactly and characterize the phases and the transitions
between them.  In section IV, we report the results of numerical
simulations in one dimension for both symmetric and asymmetric
transport of masses.  In section V, the model in one dimension is
mapped exactly to a lattice gas model whose properties are used to
deduce the steady state mass distribution exactly in two extreme
limits. In section VI, we map this lattice gas model further to an
interface model and make connections to other well studied interface
models in some limits. We study the dynamics via computing the width
of the interface both analytically in some limiting cases and
numerically otherwise. We identify two new fixed points of interface
dynamics in $(1+1)$ dimension. In section VII, we generalize our model
to include an arbitrary fragmentation kernel and obtain exact results for
special choices of this kernel via mappings to other solvable models of
statistical mechanics. In the Appendix we outline the exact solution
for uniform fragmentation kernels. We finally conclude with a summary
and a few open questions in section VIII.

\section{The Model}

Our model of diffusion, aggregation and dissociation is defined on a
lattice, and evolves in continuous time.  For simplicity we define it
on a one-dimensional lattice with periodic boundary conditions
although generalizations to higher dimensions are quite
straightforward.  Beginning with a state in which the masses are
placed randomly, a site $i$ is chosen at random. Then one of the
following events can occur:
\begin{enumerate}
\item Diffusion and Aggregation:
With rate $p_1$, the  mass $m_i$ at site $i$ moves either to
site $i-1$ or to site $i+1$.
If it moves to a site which already has some 
particles, then the total mass just adds up; thus $m_i\to 0$ and $
m_{i\pm 1}\to m_{i\pm 1}+m_i$.
\item Chipping (single-particle dissociation):
With rate $p_2$, a bit of the mass at the site
``chips'' off, {\it i.e.} provided $m_i\geq 1$, a single particle leaves
site $i$ and moves with equal probability to one of the neighbouring
sites $i-1$ and $i+1$; thus $m_i\to m_i-1$ and $m_{i\pm 1} \rightarrow
m_{i\pm 1}+1$.
\end{enumerate}
We rescale the time, $t\to p_1t$, so that the diffusion and
aggregation move occurs with rate $1$ while chipping occurs with
rate $w=p_2/p_1$.
 
This model clearly is a very simplified attempt to describe systems
with aggregation and dissociation occuring in nature. For example, if
one is thinking of gelation phenomena, then a polymer of size $k$ is
represented by a point particle of mass $k$ in our model. Thus we
ignore the spatial shape of the real polymer which however can play an
important role under certain situations. We have also assumed that the
fusion of masses after hopping or chipping occurs instantaneously,
i.e., the reaction time scale is much smaller than the
diffusion time scale. Thus our model is diffusion-limited.  A somewhat
more severe assumption is however that both desorption and diffusion
rates are independent of the mass. In a more realistic situation these
rates will depend upon the mass. However, our aim here is not to study
any specific system in full generality, but rather to identify the
mechanism of a dynamical phase transition, if any, in the simplest
possible scenario involving these microscopic processes.  If one is
interested in a more realistic description of any specific system, one
could and should include these features in the model. But for the
purpose of this paper, we stay with the simplest version and show
below that even within this simplest scenario novel, dynamical phase
transitions occur which are nontrivial yet amenable to analysis.

In this model, the total mass $M$ is conserved and fixed by the
initial condition. Let $\rho=M/N$ denote the density, i.e., mass per
site where $N$ is number of sites of the lattice. In the above
definition of the model, mass at each site can move symmetrically
either to the left or to the right with equal probability. We call
this the symmetric conserved-mass aggregation model (SCMAM).  In this
paper we also study the fully asymmetric version of the model where
masses are constrained to move only in one direction (say to the
left). We call this the asymmetric conserved-mass aggregation model
(ASCMAM).  In the sections that follow, we show that the
nonequilibrium critical behaviour of SCMAM and ASCMAM belong to
different universality classes.  This can be traced to the fact that
in the asymmetric case there is a nonzero mass current density in the
system.

In both of these models, there are only two parameters, namely the
conserved density $\rho$ and the ratio $w=p_2/p_1$ of the rate of
chipping of unit mass to that of hopping of the entire mass 
on a site, as a
whole. The question that we mainly address in this paper is: given
$(\rho, w)$, does the system reach a steady state in the long time
limit? If so, how can we characterize this steady state? We show below
that indeed for all $(\rho,w)$, the system does reach a steady
state. This is a nonequilibrium steady state in the sense that for
generic values of $(\rho, w)$ it is not described by the Gibbs
distribution associated with some Hamiltonian. In order to
characterize the steady state we study the single site mass
distribution function $P(m,t)$ as the time $t\to
\infty$. We show below that there exists a critical curve in the
$(\rho,w)$ plane across which the steady state behaviour of the system,
as characterized by $P(m)$, undergoes a novel phase transition.

Let us summarize our main results: In our model, there are two
competing dynamical processes. The diffusion cum coalescence move 
tends to produce massive aggregates at the expense of smaller masses,
and in this process, also creates more vacant sites. The chipping of
single units of mass, on the other hand, leads to a replenishment of
the lower end of the mass spectrum. The result of this competition is
that two types of steady states are possible, and there is a dynamical
phase transition between the two across a critical line $\rho_c(w)$ in
the $(\rho, w)$ plane. For a fixed $w$, if $\rho<\rho_c(w)$, the
steady state mass distribution $P(m)$ decays exponentially for large
$m$. At $\rho=\rho_c(w)$, $P(m)$ decays as a power law $P(m)\sim
m^{-\tau}$ for large $m$, where the exponent $\tau$ is the same everywhere
on the critical line $\rho_c(w)$. A more striking and interesting
behaviour occurs for $\rho>\rho_c(w)$. In this phase, $P(m)$ decays as
the same power law $\sim m^{-\tau}$ for large $m$ as at the critical
point, but in addition develops a delta function peak at
$m=\infty$. Physically this means that an infinite aggregate forms
that subsumes a finite fraction of the total mass, and coexists with
smaller finite clusters whose mass distribution has a power law tail.
In the language of sol-gel transitions, the infinite aggregate is like
the gel while the smaller clusters form the sol. However, as opposed
to models of irreversible gelation where the sol disappears in the
steady state, in our model the gel coexists with the sol, which has a
power law distribution of mass.  Interestingly, the mechanism of the
formation of the infinite aggregate in the steady state resembles
Bose-Einstein condensation (BEC), though the condensate (the infinite
aggregate here) forms in real space rather than momentum space as in
conventional BEC.

This nontrivial dynamical phase transition occurs in both the models
SCMAM and ASCMAM in all spatial dimensions $d$ including $d=1$. We
expect that the exponent $\tau$ depends on the dimension $d$ but is
universal with respect to lattice structures and initial
conditions. However, the bias in the movement of masses that
distinguishes the two models SCMAM and ASCMAM is a relevant
perturbation and the corresponding exponents $\tau_s$ (for SCMAM) and
$\tau_{as}$ (for ASCMAM) differ from each other.

We comment on the relationship of our model and results to earlier
work on related models.

\noindent (i) Takayasu and coworkers have
studied\cite{Takayasu} a lattice model where masses diffuse and
aggregate upon contact as in our model. However our model differs from
the Takayasu model in the following important way.  In that
model there is a nonzero rate of injection of a single
unit of mass from the outside into each lattice site, whereas in our model the
Takayasu injection move is replaced by the `chipping' 
of a single unit of mass to a
neighbouring lattice site. Thus in our model total mass is conserved
as opposed to the Takayasu model where total mass increases linearly
with time. In the Takayasu model the mass distribution $P(m)$ has a
power law decay in the steady state\cite{Takayasu} but there is no
phase transition as in our conserved model. Finally, while the
directed and undirected versions of the Takayasu model evolve in the
same way, in our model directionality in motion changes the
universality class.

\noindent (ii) In the context of polymer chain growth kinetics,
models of aggregation in `dry' environments have been studied earlier
only within a rate equation approach\cite{Vigil,Krapiv}.  Although the
aggregation of polymers and dissociation of single monomers were
allowed in those models, they lacked the important process of the
local diffusion of masses which is included in our lattice model.

\noindent (iii) Models of
vacancy cluster formation consider the attachment and
detachment of single vacancies to clusters \cite{Anantha}, often with rates that
depend on the cluster size. This would correspond to allowing only
chipping moves in our model (Section V A). Since cluster aggregation
moves are absent in this case, there is no tendency to form very large
clusters and the mass distribution decays exponentially with $m$.  

\noindent (iv) Recently, lattice gas models have been proposed \cite{campi}
to describe the distribution of droplets in fast-expanding systems
such as in the fragmentation process following a nuclear collision.
The distribution of fragments shows a pronounced peak at the large mass
end, reminiscent of our infinite aggregate. However, the distribution 
of the remaining fragments decays exponentially, and not as a power law
as in our case. 

\noindent (v) Bose-Einstein-like condensation in real space has also
been found in lattice gas models of traffic \cite{ferrari,evans} in
which different cars have different maximum speeds, chosen
from an {\it a priori} specified distribution. This corresponds to
phase separation into high density and low density regions, a
phenomenon which is also found when randomness in hopping rates
is associated with fixed points in space, rather than with cars
\cite{tripathy,bengrine}. An important difference
between these studies and ours is that they involve quenched disorder
(the assignment of different maximum speeds, either to cars,
or to different lattice sites), whereas the BEC phenomenon
in our case occurs in the absence of disorder, in a translationally 
invariant system.

\section{Mean Field Theory}

In this section, we study the conserved mass models within the mean
field approximation which keeps track only of the distributions of
masses, ignoring correlations in the occupancy of adjacent sites. The
mean field theory is identical for both the symmetric and asymmetric
models.  This is a defect of the MF approximation as, in fact, the
existence of a nonzero mass current in the asymmetric model affects
fluctuations in an important way. Although MF theory misses this
important aspect of the physics, it is still instructive since it reproduces
the phase diagram correctly, at least qualitatively. Besides, in high
dimensions where fluctuations are negligible, the MF theory captures
the correct physics even quantitatively.
 
Ignoring correlations between masses at neighbouring sites on the lattice, 
one can directly write down the evolution equation for $ P(m,t)$, the
probability that any site has a mass $m$ at time $t$.
\begin{eqnarray}
\frac{dP(m,t)} {dt} &=& -(1+w)[1+s(t)] P(m,t) 
+ w P(m+1,t)  \nonumber \\
&+&w s(t) P(m-1,t)+ P*P ;\;\;\; m \geq 1~~ \label{eq:mft1}\\
\frac{dP (0,t)} {dt} &=& - (1+w)s(t) P(0,t) + wP(1,t) + 
s(t) \label{eq:mft2}. 
\end{eqnarray} 
Here $ s(t) \equiv 1-P(0,t)$ is the probability that a site is
occupied by a mass and 
$P*P=\sum_{m^{\prime}=1}^{m}P(m^{\prime},t)P(m-m^{\prime},t)$ is a
convolution term that describes the coalescence of two masses.

The above equations enumerate all possible ways in which the mass at a
site might change. The first term in Eq. (\ref{eq:mft1}) is the
``loss'' term that accounts for the probability that a mass $m$ might
move as a whole or chip off to either of the neighbouring sites, or that
a mass from a neighbouring site might move or chip off to the site in
consideration. The probability of occupation of the neighbouring site,
$s(t) = \sum_{m=1} P(m,t)$, multiplies $P(m,t)$ within the mean-field
approximation where one neglects the spatial correlations in the
occupation probabilities of neighbouring sites. The remaining three
terms in Eq. (\ref{eq:mft1}) are the ``gain'' terms enumerating the
number of ways that a site with mass $m^{\prime} \neq m$ can gain the
deficit mass $m -m^{\prime}$. The second equation Eq. (\ref{eq:mft2})
is a similar enumeration of the possibilities for loss and gain of
empty sites.  Evidently, the MF equations conserve the total mass.

To solve the equations, we compute the generating function, $Q(z,t) =
\sum_{m=1}^{\infty} P(m,t)z^{m}$ from Eq. (\ref{eq:mft1}) and set $
\partial Q / \partial t =0$ in the steady state. We also need to use
Eq. (\ref {eq:mft2}) to write $P(1,t)$ in terms of $s(t)$. This gives
us a quadratic equation for $Q$ in the steady state.  Choosing the
root that corresponds to $Q(z=0) = 0$, we find
\\
\begin{eqnarray}
Q(z) &=&{{w+2s+ws}\over {2}}-{w\over {2z}}-{wsz\over {2}} 
\nonumber \\
&+& ws{(1-z)\over {2z}}\sqrt {(z-z_1)(z-z_2)}.  \label{eq:qsol}
\end{eqnarray}
where $z_{1,2}=(w+2\mp 2\sqrt {w+1})/ws$.
The value of the occupation probability $s$ is
fixed by mass conservation which implies that $\sum mP(m)=M/L\equiv \rho$.
Putting ${\partial}_zQ (z=1)=\rho$, the resulting relation between $\rho$ 
and $s$ is \begin{equation}
2\rho = w(1-s) - ws\sqrt {(z_1-1)(z_2-1)}~.
\label{eq:defq}
\end{equation}

The steady state probability distribution $P(m)$ is the coefficient of 
$z^m$ in $Q(z)$ and can be obtained from $Q(z)$ in Eq. (\ref{eq:qsol})
by evaluating the integral 
\begin{equation}
P(m) = {1\over {2\pi i}}\int_{C_o} \frac {Q(z)} {z^{ m+1}} dz
\label{eq:contour}
\end{equation}
over the contour $C_o$ encircling the origin. 
The singularities of the integrand govern the asymptotic behaviour of 
$P(m)$ for large $m$. Clearly the integrand has branch cuts at 
$z=z_{1,2}$. For fixed $w$, if one increases the density $\rho$, the
occupation probability $s$ also increases as evident from Eq. 
(\ref{eq:defq}).
As a result, both the roots $z_{1,2}$ start decreasing. As long as the
lower root $z_1$ is greater than $1$, Eq. (\ref{eq:defq}) is well defined
and the analysis of the contour integration around the branch cut
$z=z_1$, yields for large $m$,
\begin{equation}
P(m) \sim e^{-m/{m^{*}}}/m^{3/2} ~,
\end{equation}
where the characteristic mass, $m^{*}=1/{\log (z_1)}$. It diverges as
$\sim (s_c-s)^{-1}$ as $s$ approaches $s_c =(w+2-2\sqrt {w+1})/w$. 
$s_c$ is the critical value of $s$ at which $z_1=1$. This exponentially 
decaying mass distribution is the signature of the ``disordered" 
phase which 
occurs for $s<s_c$ or equivalently from Eq. (\ref{eq:defq}) for
$\rho < {\rho}_c(w)={\sqrt {w+1}} -1$.

When $\rho={\rho}_c$, we have $z_1=1$, and analysis of the contour around 
$z=z_1=1$ yields a power law decay of $P(m)$,
\begin{equation}
P(m)\sim m^{-5/2}.
\end{equation}
As $\rho$ is increased further beyond 
${\rho}_c(w)$, $s$ cannot increase
any more because if it does so, the root $z_1$ would be less than $1$ 
(while the other root $z_2$ is still bigger than $1$) and the right
hand side of 
Eq. (\ref{eq:defq}) would become complex. The only possibility is that $s$
sticks to its critical value $s_c$ or equivalently the lower root $z_1$
sticks to $1$. Physically this implies that adding more particles   
does not change the occupation probability of sites. This can happen only 
if all the additional particles (as $\rho$ is increased) aggregate on a 
vanishing fraction of sites, thus
not contributing to the occupation of the others. Hence in this 
``infinite-aggregate" 
phase $P(m)$ has an infinite-mass aggregate, in addition to a cluster
distribution with a power law
decay $m^{-5/2}$. Concomitantly Eq. (\ref{eq:defq}) ceases to hold, and 
the relation now becomes 
\begin{equation}
\rho = {w\over {2}}(1-s_c) + \rho_{\infty}
\end{equation}
where $\rho_{\infty}$ is the fraction of the mass in the infinite aggregate.
The mechanism of the formation of the aggregate is reminiscent of Bose 
Einstein condensation. In that case, for temperatures in which a macroscopic
condensate exists, particles added to the system do not contribute to the 
occupation of the excited states; they only add to the condensate, as 
they do to the infinite aggregate here.

Thus the MF phase diagram (see inset of Fig. 1) of the system consists
of two phases, ``Exponential" and ``Aggregate", which are separated by
the phase boundary, $\rho_c(w)={\sqrt {w+1}}-1$.  We note that in
conserved-aggregation models in `dry' environment studied earlier
within a rate equation approach\cite{Vigil,Krapiv}, the steady state
mass distribution also changed from an exponential distribution to a
power law as the density was increased to a critical value. However,
the existence of the infinite aggregate in the steady state for
$\rho>{\rho}_c(w)$ was not identified earlier.

\section{Numerical Simulation in One Dimension}

In order to see if the MF phase diagram remains at least qualitatively
correct in lower dimensions, we have studied both SCMAM and  
ASCMAM
using Monte Carlo simulations on a one-dimensional lattice with
periodic boundary conditions. Although we present results here for a
relatively small size lattice, $N=1024$, we have checked our results
for larger sizes as well. We confirmed that all the qualitative
predictions of the mean-field theory remain true in $1$-d though the
exponents change from their MF values.

Figure 1 displays two numerically obtained plots of $P(m)$ in the
steady state of SCMAM. For fixed $w=1.0$, we have measured $P(m)$ for
two values of the density, namely $\rho=0.2$ and $\rho=3.0$. For
$\rho=0.2$, we find exponential decay of $P(m)$ (denoted by $\times$
in Fig. 1).  For $\rho=3.0$, $P(m)$ decays as a power law (denoted by
$+$ in Fig.  1) which is cut off by finite size effects but in
addition, there is a sharp peak at a much larger mass signalling the
existence of the `infinite' aggregate as predicted in the MF theory.
We confirmed that the mass $M_{agg}$ in this aggregate grows linearly
with the size, and that the spread $\delta M_{agg}$ grows sublinearly,
implying that the ratio $\delta M_{agg}/M_{agg}$ approaches zero in
the thermodynamic limit. As one decreases $\rho$ for fixed $w$, the
mass $M_{agg}$ decreases and finally vanishes at the critical point,
$\rho_c(w)$ where $P(m)$ only has a power law tail. For $w=1$, we find
numerically $\rho_c(1)\approx 0.39$. In the inset of Fig. 1, we plot
the numerical phase boundary in the $(\rho,w)$ plane (denoted by
closed circles). This is only a rough estimate of the phase boundary
obtained from small lattice sizes and the numerical accuracy of these
points should not be taken too seriously.  For comparision, we also
plot the MF phase boundary $\rho_c(w)={\sqrt {(w+1)}}-1$.

In Fig. 2 and in its inset, we present similar plots for the
asymmetric model ASCMAM. Here the steady state mass distribution
function $P(m)$ is plotted for two values of the density $\rho=0.2$
and $\rho=10.0$ with fixed $w=1.0$.  In the first case, $P(m)$ decays
exponentially whereas in the second case it has a power law tail and
in addition the `infinite' aggregate as in the case of SCMAM.

The exponent $\tau_{s}$ for SCMAM which characterizes the finite-mass
fragment power law decay for $\rho>{\rho}_c(w)$ is numerically found
to be $ 2.33\pm .02$ and remains the same at the critical point
$\rho=\rho_c(w)$. In the asymmetric model ASCMAM, we find the
corresponding exponent $\tau_{as}\approx 2.05$ within numerical error.
Note however that because the total mass and hence the mass density,
$\rho=\sum mP(m)$ is conserved and finite, the decay of $P(m)$ must be
faster than $m^{-2}$ for large $m$ to avoid ultraviolet divergence. In
ASCMAM the numerical value of $\tau_{as}$ in $1$-d is very close to
$2$ suggesting perhaps that $P(m)$ decays as $m^{-2}$ with additional
logarithmic corrections such that $\rho$ remains finite. But within
our simulations, it is not easy to detect these additional logarithmic
factors. Thus clearly in $1$-d, SCMAM and ASCMAM belong to different
universality classes.


\section{Mapping to a Lattice Gas Model in One Dimension}

In this section, we show that in one dimension the mass model studied
above can be mapped exactly onto a lattice gas (LG) model consisting
of particles and holes. In the language of the LG model, it is
somewhat easier to understand the two phases and the transition
between them. Besides, for certain limiting values of the parameters,
the steady state of the LG model can be solved exactly. This then
provides, via the mapping, exact solutions for the mass model in those
limiting cases.

Consider the mass model (both SCMAM and ASCMAM) on a ring $R$ of $N$
lattice sites. Let $m_i$ denote the mass at site $i$ of $R$ in a given
configuration. Let $M=\sum_{i=1}^{N} m_i$ denote the total mass on
$R$.  We first construct a new ring $R'$ consisting of $L=N+M$
sites. For every lattice site $i$ of ring $R$, we put a particle
(labelled by $i$) on ring $R'$ such that the $i$-th and $(i+1)$-th
particle on $R'$ are separated exactly by $m_i$ holes. The ring $R'$
will therefore have $N$ particles and $M$ holes. Also by construction,
these particles on $R'$ are hard core, i.e., any site of $R'$ can
contain at the most one particle. Thus every mass configuration on $R$
maps onto a unique particle-hole configuration on $R'$. In Fig. 3, we
give an example of this mapping. We also note that particle density
${\rho}'$ on $R'$ is simply related to the mass density $\rho$ on $R$
via, ${\rho}'=N/(N+M)=1/(1+\rho)$.

Given this exact mapping of a mass configuration on $R$ to a particle
hole configuration on $R'$, we now examine the correspondence between
the mass dynamics on $R$ and the particle dynamics on $R'$. Consider a
pair of neighbouring sites $(i-1)$ and $i$ on $R$ where the masses are
$m_{i-1}$ and $m_i$. This translates to having $m_{i-1}$ holes to the
right of $(i-1)$-th particle and $m_i$ holes to the right of $i$-th
particle on $R'$. Consider first the chipping move that occurs with
rate $p_2$ (Fig. 3). 
If a single unit of mass chips 
off the $(i-1)$-th site and
moves to $i$-th site (i.e., $m_{i-1}\to m_{i-1}-1$ and $m_i\to
m_{i}+1$) on $R$, it corresponds to the $i$-th particle on $R'$
hopping to its neighbouring site to the left with rate
$p_2$. Similarly, the reverse move of a unit mass chipping off $i$-th
site to $(i-1)$-th site on $R$ (i.e., $m_{i}\to m_{i}-1$ and
$m_{i-1}\to m_{i-1}+1$) corresponds to the $i$-th particle on $R'$
hopping one step to the right with rate $p_2$. Thus the chipping move
of SCMAM on $R$ corresponds precisely to the ``symmetric exclusion
process" (SEP)\cite{Lig} on $R'$ where a particle can hop to its
nearest neighbour on either side provided the neighbour is
empty. Similarly the chipping move of ASCMAM on $R$ corresponds
exactly to the ``asymmetric exclusion process"
(ASEP)\cite{Lig,Krug1,SZ} on $R'$ where particles move only along one
direction on $R'$.
  
But in addition to chipping, we also have the diffusion and
aggregation move in the mass model that occurs with rate
$p_1$ (as shown for the block of particles in Fg. 3).
Consider once again a pair of sites $(i-1)$ and $i$ with
respective masses $m_{i-1}$ and $m_{i}$ on $R$. Suppose the whole mass
$m_{i-1}$ moves to the $i$-th site (i.e., $m_{i-1}\to 0$ and $m_{i}\to
m_{i}+m_{i-1}$). This would mean that on $R'$, the $i$-th particle
jumps to the farthest available hole (without crossing the $(i-1)$-th
particle) to its left with rate $p_1$. Similarly the reverse move,
$m_{i}\to 0$ and $m_{i-1}\to m_{i-1}+m_{i}$, would translate on $R'$
to the $i$-th particle jumping to the farthest available hole to its
right (without crossing the $(i+1)$-th particle) with rate $p_1$. In
the asymmetric version of the model, the particle can jump to the
farthest available hole (without crossing the next particle) in one
direction only.

To summarize, in our LG model we have a bunch of hard core particles
with particle density ${\rho}'$. The world lines of particles can not
touch or cross each other due to their hard core nature. There are two
possible moves for each particle. With rate $p_2$, a particle moves to
its adjacent site (if it is empty) and with rate $p_1$ the particle
jumps to the farthest available hole maintaining the hard core
constraint.  In SCMAM, the particle can hop with equal probability to
the left or right. In ASCMAM, it hops only in one direction, say to
the right. In this LG language, it is the competition between the
short range and long range hopping of the particles that is
responsible for the phase transition. As in the mass model, the only
two parameters of the model are the ratio of the two rates,
$w=p_2/p_1$ and the density of particles ${\rho}'$. We also note that
$P(m)$ in the mass model corresponds to the size distribution of hole
clusters in the LG model.

In the following we fix the density of particles ${\rho}'$ and study
the two extreme limits $w\to \infty$ and $w=0$ where exact results
can be obtained.

\subsection{Only chipping : $w\to \infty$}

Since $w=p_2/p_1$, the limit $w\to \infty$ corresponds to $p_1=0$ with
$p_2$ remaining nonzero. This means only chipping moves are allowed in
the mass model. As mentioned in Section II, in this limit the model
has some resemblance to models of vacancy cluster formation
\cite{Anantha}, with mass
in our model representing the number of vacancies in a cluster. Chipping
off from clusters with $m \geq 2$ corresponds to vacancy detachment,
whereas the chipping move for $m=1$ is tantamount to either diffusive hopping
(if the neighbouring site is empty), or to attachment to a cluster
(if the neighbouring site has a mass on it).

In the LG version, the `only-chipping' model  corresponds to just the
exclusion process\cite{Lig,Krug1} either symmetric (SEP)
(corresponding to SCMAM) or asymmetric (ASEP) (corresponding to
ASCMAM). Both the SEP and ASEP have been studied in great
detail and several exact results are known\cite{Lig,Krug1}. In
particular it is known that in the steady states of both SEP and ASEP,
all configurations are equally likely\cite{Lig,Krug1}. This means that
in the thermodynamic limit, the steady state has a simple product
measure: the probability $P(m)$ of having exactly $m$ holes following
a particle is simply given by, $P(m)= {\rho}'(1-{\rho}')^m$.  Using
${\rho}'=1/(1+\rho)$, we thus obtain an exact result for $P(m)$ of the
mass model in the $w\to \infty$ limit,
\bea
P(m)= {{\rho^m}\over {(1+\rho)^{m+1}}}
\label{eq:pm9}
\eea
which clearly demonstrates the exponential decay of $P(m)$ for large $m$.

We note in passing that Eq.(\ref{eq:pm9}) also describes the
distribution of masses in any dimension in an `only chipping' model in
which the rate of chipping at site $i$ is proportional to the mass
$m_i$ on that site. In this case, evidently every unit of mass
performs a simple random walk, and the problem is tantamount to that
of $M$ independent random walkers on the lattice. Standard methods of
statistical mechanics can then be used to describe the steady
state. The grand partition function is then 
\bea
Z = (1 + z + z^2 \cdots)^N = {1 \over (1-z)^N}
\label{eq:Grand}
\eea
where $z$ is the fugacity, and the probability of finding $m$ particles on
a site is
\bea
P(m) = (1-z) z^m .
\label{eq:Fugacity}
\eea
Eliminating $z$ in favour of $\rho = z/(1-z)$, we see that
Eq.(\ref{eq:Fugacity}) reduces to Eq.(\ref{eq:pm9}).   

\subsection{No chipping: $w=0$}

We now consider the other limit $w=0$ where there is no chipping and
the masses only diffuse as a whole and aggregate with each other. In the
LG language, this would mean that particles undergo only long
range hopping to the farthest available hole (without crossing the next
particle). 
 
In order to find the exact steady state in this limit, we first
consider the mass model on a finite ring $R$ of $N$ sites and total
mass $M=\rho N$. As time progresses, masses diffuse and coagulate with
each other and the number of empty sites decreases. Since the
diffusive motion confined in a finite region of space is ergodic,
eventually all the masses will coagulate with each other and there
will be precisely one single big conglomerate with mass $M$ which will
move on the ring. Thus in the steady state, one gets exactly,
$P(m)=\delta (m-\rho N)$. In the LG language, this would mean that on
a finite lattice, the particles and holes will become completely phase
separated in the steady state.

It is also useful to study the approach to this phase separated steady
state. We first note that on an infinite $1$-d lattice, the time
dependent single site mass distribution function, $P(m,t)$ can be
exactly solved in the $w=0$ limit\cite{Majhuse}. It was shown exactly
in Ref. \cite{Majhuse} that in the scaling limit, $m\to \infty$, $t\to
\infty$ but keeping $m/{\sqrt {t}}$ fixed, the function $P(m,t)\sim
t^{-1/2}S(m/c{\sqrt {t}})$ where the constant $c$ depends on the initial
condition but the scaling function $S(x)$ is universal and is given by
\bea
S(x)= {{\pi x}\over {2}}\exp \big [{-{{\pi x^2}\over {4}}}\big ].
\eea
In the LG model, this would mean that as time progresses, the system
undergoes phase separation and breaks into domains of particles and
holes. The average linear size of these domains grows with time as
$l(t)\sim t^{1/2}$ at late times. At this point it is useful also to
note that in usual models of coarsening with locally conserved
dynamics, the domain size grows as $l(t)\sim t^{1/3}$\cite{Bray} as
opposed to $t^{1/2}$ here.  This, however, is not entirely surprising
since in our model, even though the particle number is conserved
globally, the long range hopping effectively reduces this to a locally
nonconserved model with growth law $t^{1/2}$.  Similar behaviour was
noticed earlier in other models of coarsening with globally conserved
dynamics\cite{Sima}. In an infinite system the domain size keeps
growing indefinitely as $l(t)\sim t^{1/2}$.  However in a finite
lattice of sites $L$, when $l(t)\sim L$ after a time $t\sim L^2$, the
domains stop growing and the system eventually breaks up into two
domains only, one of particles and the other of holes.  We note that
all the conclusions made in this subsection are equally valid for both
symmetric and asymmetric versions of the model.
 
To summarize this section, we find that for fixed $\rho$ in one
dimension, the two extreme limits $w\to 0$ and $w\to \infty$ are
exactly solvable for their steady states. In the limit $w\to 0$, the
system has a phase separated steady state in the LG model which in the
mass model corresponds to having a single massive aggregate. In the
other limit $w\to \infty$, the LG corresponds to simple exclusion
process with product measure steady state, which corresponds to an
exponential mass distribution in the mass model.  Thus there is a
competition between the long range hopping that tends to create phase
separation and the short range hopping that tends to mix the particles
and holes to produce a product measure steady state.  As $w$ is
increased from $0$ for fixed $\rho$, the massive aggregate coexists
with power law distributed smaller masses (or hole clusters in the LG
language) up to some critical value $w_c(\rho)$. For $w>w_c(\rho)$, the
massive aggregate disappears and the cluster size distribution of
holes becomes exponential.

The phase separated state found in the lattice model is reminiscent 
of the state found in lattice gas models of traffic with quenched
disorder, either in the distribution of maximum velocities of different
cars \cite{ferrari,evans} or at different locations on the road 
\cite{tripathy,bengrine}. Our model, by contrast, has no disorder and is
translationally invariant to start with, although this symmetry is
broken in the phase-separated state.

\section{Dynamics in one dimension: Mapping to Interface Models in
$(1+1)$ dimension}

In the previous sections we have studied the static properties of the
model in the steady state. However it is also important to study the
dynamics of the model in the steady state. The universal features of
the dynamics is usually best captured by time dependent correlation
functions in the steady state. In this section we however take a
slightly different route. Instead of studying the time-dependent
correlation functions directly in the mass or the equivalent latice
gas model, we first map the LG model onto an interface model and then
study the time dependent properties of the width of the fluctuating
interfaces. The advantage of this route is that not only does it
capture the essential universal features of the dynamics, but it also
makes contact with other well studied interface models in certain
limiting cases.

There is a standard way\cite{Krug1} to map a LG configuration in one
dimension to that of an interface configuration on a $1$-d
substrate. One defines a new set of variables $\{S_i\}$ such that
$S_i=1$ if the $i$-th site is occupied by a particle and $S_i=-1$ if
it is empty. Then the interface height $h_i$ at site $i$ is defined
as, $h_i=\sum_{j=1}^{i} S_j$. Thus the overall tilt of the interface,
$\tan {\theta}= (h_L-h_1)/L=2{\rho}'-1$ is set by the particle density
${\rho}'$ in the LG model.

The different dynamical moves in the LG model can be translated in a
one to one fashion to corresponding moves of the interface. For
example, if a particle at site $i$ jumps one unit to the right, it
corresponds to the decrease of height $h_i$ by $2$ units, $h_i\to
h_i-2$. Similarly, if the particle at site $(i+1)$ jumps one unit to
the left, the height $h_i$ increases by $2$ units, $h_i\to
h_i+2$. Thus the nearest neighbour hopping of particles to the left or
right in the LG model corresponds respectively to deposition and
evaporation moves in the interface model. Similarly the long range
hopping of a particle to the farthest available hole before the next
particle would translate to nonlocal moves in the interface as shown
in Fig. 3.  Once again, the ratio of the rates of
evaporation-deposition moves to that of the nonlocal moves is given by
the parameter $w$.

A natural measure of the fluctuations of the interface is its width
defined for a finite system of size $L$ as,
\bea
W(L,t)=\sqrt {{1\over {L}}\sum_{i=1}^{L} {\big [ h_i-{\bar h} \big ]}^2}
\label{width}
\eea
where ${\bar h}={\sum_{i=1}^{L} h_i}/L$. The width $W(L,t)$ is expected to
have a scaling form\cite{Krug1},
\bea
W(L,t)\sim L^{\chi}F\big ( t/L^z\big ),
\eea
in the scaling limit with large $L$, large $t$ but keeping $t/L^z$
finite.  The exponents $\chi$ and $z$ are respectively the roughness
and the dynamical exponents and the scaling function $F(x)$ is
universal with the aymptotic behaviour: $F(x)\to O(1)$ as $x\to
\infty$ and $F(x)\sim x^{\beta}$ as $x\to 0$ where $\beta=\chi z$.
The exponents $\chi$ and $z$ characterize the universality classes of
the interface models.

In the following we keep the particle density ${\rho}'$ fixed and
investigate the width of the corresponding interface model $W(L,t)$ by
varying the parameter $w$. How do the exponents $\chi$ and $z$ that
characterize the universal behaviour of interface fluctuations change
as one varies $w$ from $0$ to $\infty$?  We study this question for
both the symmetric and the asymmetric models and denote the respective
exponents by $({\chi_w}^{(s)}, {z_w}^{(s)})$ and $({\chi_w}^{(as)},
{z_w}^{(as)})$.  In the following three subsections, we study the
width $W(L,t)$ in both models for three special values of $w$, namely
$w\to \infty$, $w=0$ and $w=w_c(\rho')$. In the first two cases we
find analytical results whereas in the last case, we present numerical
results only.

\subsection{Only chipping: $w\to \infty$}

We first consider the symmetric mass model SCMAM whose lattice gas
equivalent corresponds to the SEP in the $w\to \infty$ limit as noted
in section V. This means that in the corresponding interface model,
nonlocal moves are absent and only the allowed moves are evaporation and
deposition subject to certain local constraints.  An important
property of this symmetric model is that the average velocity of the
interface is zero as the evaporation and deposition occurs with equal
probability.  It is well known\cite{Krug1} that the continuum version
of this discrete interface model corresponding to SEP is well
described by the Edwards-Wilkinson equation\cite{EW,Hamm}. This linear
evolution equation can be easily solved and one gets the exact
exponents\cite{EW,Hamm},
\bea
{\chi_{\infty}}^{(s)}={1\over {2}};\quad\quad\quad\quad
{z_{\infty}}^{(s)}=2.
\label{edw}
\eea

The LG equivalent of the asymmetric mass model ASCMAM in the $w\to
\infty$ limit is the ASEP. In the corresponding interface model the
allowed moves are either only deposition or only evaporation (but not
both) depending upon the direction of particle motion in ASEP. Thus
the interface has a nonzero average velocity and its continuum version
is known\cite{Krug1} to be described by the nonlinear KPZ
equation\cite{KPZ}. In one dimension, the KPZ equation can be solved
exactly and the exponents are known to be\cite{KPZ},
\bea 
{\chi_{\infty}}^{(as)}={1\over {2}};\quad\quad\quad\quad
{z_{\infty}}^{(as)}={3\over {2}}.
\label{kpz}
\eea

Thus in the limit $w\to \infty$, our model reduces to two known
interface models for symmetric and asymmetric cases respectively and
the corresponding exponents are obtained exactly.

\subsection{No chipping: $w=0$}

We have mentioned in section V that in the $w\to 0$ limit, as time
progresses the LG phase separates into domains of particles and
holes. In the equivalent spin representation where a particle is
represented by an up spin, $S_i=1$ and a hole by a down spin,
$S_i=-1$, this then represents a spin model coarsening with time with
average domain size growing as, $l(t) \sim t^{1/2}$. In a finite
system of length $L$, eventually the system breaks up into two domains
of opposite signs. In the interface representation, this would mean
that the system would develop a single mound. Mound formation has also
been studied recently in other interface models\cite{Krug2}.

We note from Eq. \ref{width} and the definition,
$h_i=\sum_{j=1}^{i}S_j$ that the calculation of the width $W(L,t)$
requires the expression for the equal time spin correlation function,
$\langle S(i,t)S(j,t)\rangle$ in the equivalent spin model. From the
general theory of coarsening it is known\cite{Bray} that for $l(t)
<<L$, the equal time spin correlation function satisfies the scaling
behaviour, $\langle S(0,t)S(r,t)\rangle \sim G(r/l(t))$. We have
assumed that the system size $L$ is large so that translational
invariance holds in the bulk of the system. Using this scaling form in
Eq.
\ref{width} and the result, $l(t) \sim t^{1/2}$, a simple
power counting gives, $W(L,t)\sim L F(t/L^2)$. Thus for $w=0$, we
have the exponents,
\bea
\chi_0=1; \quad\quad\quad\quad z_0=2,
\eea
for both the symmetric and asymmetric models. The roughness exponent 
$\chi_0=1$
can be easily understood from the fact that in the steady state the
system develops a single mound whose maximum height is of $O(L)$ and
hence the width of the fluctuations is also of order $O(L)$.    

\subsection{Critical Point: $w=w_c(\rho)$}
In this subsection, we keep the density ${\rho}'=1/(1+\rho)$ fixed
and tune $w$ to its critical value, $w=w_c(\rho)$ and calculate
the width $W(L,t)$ of the interface. Unfortunately we are unable to 
obtain any analytical result for this critical case and will only
present numerical estimates.

We first consider the symmetric model. In this case we fix $\rho=1$.
In the LG language this means particle density, $\rho'=1/2$.  We first
estimate the critical point, $w_c(\rho=1)\approx 3.35$ by simulating
the equivalent mass model. Next we fix the value of $\rho=1$ and
$w=3.35$ in the interface model and measure the width $W(L,t)$ for
different lattice sizes, $L=16$, $32$, $64$, $128$ and $256$. In order
to verify the scaling form, $W(L,t)\sim L^{\chi_c}F(t/L^{z_c})$, we
plot in Fig. 4, $W/L^{\chi_c}$ as a function of $t/L^{z_c}$ for
different $L$.  The best collapse of data is obtained with the choice,
${\chi_c}^{(s)}\approx 0.67$ and ${z_c}^{(s)}\approx 2.1$.

The asymmetric case however is much more complicated due to the
possible existence of logarithmic factors in one dimension and perhaps
also due to other strong corrections to scaling.  Proceeding as in the
symmetric case, we first estimate the critical point,
$w_c(\rho=1)\approx 0.77$ and the measure the width $W(L,t)$. In this
case we did not find a good data collapse using the canonical scaling
form, $W(L,t)\sim L^{\chi}F(t/L^{z})$.  Instead we tried collapsing
the data assuming, $W(L,t)\sim W_{st}F(t/L^z)$ where $W_{st}$ is the
steady state width $W(L,\infty)$. This gives a somewhat better
convergence to collapse as $L$ increases as shown in Fig. 5. with the
choice $z\approx 1.67$.  In the inset, we plot $W_{st}$ vs. $L$. It is
difficult to estimate the asymptotic growth $W_{st}\sim L^{\chi}$
(possibly with a logarithmic corrections) from the available data. A
naive linear fit to the log-log plot of $W_{st}$ vs. $L$ gives an
estimate of the slope, $\chi\approx 0.68$. Thus given the available
data, we find the approximate estimates for the exponents,
${{\chi}_c}^{(as)}\approx 0.68$ and ${z_c}^{(as)}\approx 1.67$ at the
critical point of the asymmetric model. However these are just
approximate estimates and one needs larger scale simulations to
determine the exponents more accurately for the asymmetric model.

\subsection{Flows}
We have also studied the width $W(L,t)$ numerically for other values
of $w$. This includes the numerical verification of analytical
predictions for the exponents in the limit of small and large $w$. We
do not present here all these details but summarize the main picture
that emerges from these studies by means of the schematic flow diagram
shown in Fig. 6. 

We find three different sets of
exponents $(\chi,z)$ that characterize the behaviour in three regions on
the $w$ axis for fixed ${\rho}$: subcritical when $w<w_c(\rho)$,
critical when $w=w_c(\rho)$ and supercritical when $w>w_c(\rho)$. The
subcritical regime is controlled by the aggregation fixed point (denoted
$AGG$ in Fig. 6) at $w=0$, i.e.,
the phase separation fixed point with $\chi=1$ and $z=2$ for both
symmetric and asymmetric models. The supercritical regime is
controlled by the fixed point at $w\to \infty$. For the symmetric
case, this is the Hammersley-Edwards-Wilkinson fixed point $HEW$
($\chi=1/2$, $z=2$),
whereas in the asymmetric case this is the KPZ fixed point
($\chi=1/2$, $z=3/2$). The fixed points $SC$ and $ASC$ in Fig. 6
 correspond to criticality
in the symmetric and asymmetric  conserved mass models respectively;
 these are new unstable
fixed points with exponents, ($\chi\approx 0.67$, $z\approx 2.1$) for
the symmetric case and ($\chi\approx 0.68$, $z\approx 1.67$) for the
asymmetric case respectively. 

\section{Generalization to arbitrary fragmentation kernel: Relation to
other models}  

In the mass model discussed so far we have considered only two
possible moves, namely ``chipping" of a single unit of mass to a
neighbouring site with rate $w$ or hopping of the mass as a whole to a
neighbouring site with rate $1$. However in a more general setting,
$k$ units of mass can break off a mass $m$ and hop to a neighbouring
site with rate $p(k|m)$ where $k\leq m$. In the equivalent lattice gas
version in one dimension, this would mean the hopping of a particle with
rate $p(k|m)$ to the $k$-th hole to the right or left without crossing
the next particle which is located at a distance $m+1$.  In the
asymmetric version, the particles jump only along one direction as
usual (either left or right).  In the model studied so far,
\bea 
p(k|m)= w\delta_{k,1}+ \delta_{k,m}
\label{ker1}
\eea
where $\delta_{i,j}$ is the Kronecker delta function. With this
mass fragmentation kernel, we have seen that the in the steady state the
system undergoes a nonequilibrium phase transition as the parameter $w$
that controls the relative strength of the two delta peaks is
varied.

The question naturally arises as to whether this phase transition
exists for arbitrary fragmentation kernel $p(k|m)$. In general, it is
hard to find the steady state analytically for arbitrary
$p(k|m)$. However, for some special choices of the kernels, it is
possible to obtain exact steady states via mapping to some other
solvable models of statistical mechanics. We list below a few of them.

In the context of traffic models, Klauck and Schadschneider recently
studied\cite{KS} an asymmetric exclusion process in one dimension
where a particle can jump either to the neighbouring hole to the right
with rate $p_1$ or to the second hole to the right with rate $p_2$.
The corresponding jump kernel can be written as
\bea
p(k|m)= p_1\delta_{k,1} + p_2\delta_{k,2}.
\eea
By generalizing the matrix product ansatz used for ASEP\cite{DE}, the
steady state of this model was shown\cite{KS} to have simple product
measure for all $p_1$ and $p_2$. Using this result, it is easy to show
that in the corresponding mass model with mass density $\rho$, the steady
state single site mass distribution $P(m)$ (same as
the probability of having a hole cluster of size $m$ in the LG
model) is simply given by, $P(m)=\rho/(1+\rho)^{m+1}$ and therefore decays 
exponentially for large $m$ for arbitrary $p_1$ and $p_2$.

Recently Rajesh and Dhar studied\cite{RD} an anisotropic directed
percolation model in $3$ dimensions. Their model can be reduced to
an asymmetric hard core LG model with the following jump kernel,
\bea
p(k|m)=p^{1-{\delta}_{k,0}}(1-p)^{m-k},
\eea
where $0\leq p \leq 1$. By mapping it to the five vertex model, the steady
state of this model was shown exactly to have a simple product measure
for all $p$. This then immediately gives for the mass model, once again, 
$P(m)=\rho/(1+\rho)^{m+1}$.

Another exact result can be derived for the following asymmetric 
mass model. Instead of discrete mass, we now consider continuum masses
at each site. The mass $m_i$ at each site $i$ evolves in discrete
time according to the following stochastic equation,
\bea
m_i(t+1)= q_{i-1,i}m_{i-1}(t) + (1-q_{i,i+1})m_i(t)
\eea
where the random variable $q_{i-1,i}$ represents the fraction of mass
that breaks off from site $(i-1)$ and moves to site $i$. We assume that
each of these fractions are independent and identically distributed in
the interval $[0,1]$ with some distribution function $\eta (q)$. We
show in the Appendix that for the special case of uniform
distribution, i.e., $\eta (q)=1$ for all $q$ in $[0,1]$, the exact
steady state distribution $P(m)$ is given by,
\bea
P(m)={4m\over {\rho^2}} e^{-2m/\rho}
\label{exact}
\eea   
where $\rho=\int_0^{\infty}mP(m)dm$ is the conserved mass density
fixed by the initial condition. In the
LG language, this would mean that if $m_i$ is the distance
between $i$-th and $(i+1)$-th particle, then the $i$-th particle
can jump to any distance between $k$ and $k+dk$ to the right (without
crossing the next particle on the right) with uniform rate
$p(k|m_i)dk={1\over {m_i}}dk$. In order to verify the exact formula 
for $P(m)$ in Eq. \ref{exact} we performed numerical simulation
in $1$-d. In Fig. 7, we show for $\rho=1$, the perfect agreement between
the theoretical cumulative mass
distribution, $F(m)=\int_0^m P(m')dm'=1-e^{-2m}-2me^{-2m}$ 
and the numerically obtained points from the simulation.
We note that a similar result was recently derived \cite{garcia} in
the context of the generalized Hammersley process \cite{hammersley}.

Thus in all these cases studied in this section, we find exponential
decay of the mass distribution for large mass in the steady state.  It
would therefore seem that only for the special kernel, $p(k|m)=
w\delta_{k,1}+ \delta_{k,m}$, is there a nontrivial phase transition
from a phase where $P(m)$ decays exponentially to another where it has
a power law decay in addition to an infinite aggregate. The presence
of two delta function peaks in $p(k|m)$ seem to be crucially
responsible for this phase transition. We believe that the phase
transition will still persist if one allows for a nonzero width to the
delta peaks at the two ends of the kernel near $k=0$ and $k=m$ but
making sure that the widths remain finite even as $m\to \infty$.  This
however needs further studies to be confirmed.
 
\section{Conclusion}

In this paper we have studied a lattice model of aggregation and
dissociation where a mass from a site can either move as a whole to a
neighbouring site with rate $p_1$ or can chip off a unit mass to a
neighbour with rate $p_2$. The hopped mass then aggregates
instantaneously with the mass that is already present at the
neighbour. The ratio of the rates, $w=p_2/p_1$ and the conserved mass
density $\rho$ are the only two parameters of the model. The steady
state of the model undergoes a phase transition as the parameters
$(\rho, w)$ are varied. In the $(\rho,w)$ plane there is a critical
line that separates two phases: (i) the `Exponential' phase where the
single site mass distribution $P(m)$ decays exponentially for large
$m$ and (ii) the `Aggregate' phase where $P(m)$ has a power law decay
in addition to a delta function peak at $m=\infty$ signifying the
presence of an infinite aggregate. On the critical line, the aggregate
vanishes but $P(m)\sim m^{-\tau}$ retains the same power law tail. We
have also studied how the universality class of this dynamical phase
transition changes on applying a bias in a particular direction of the
mass transport.

We have solved the model exactly within the mean field theory and
presented numerical results for one dimension. Besides, by exploiting
a mapping to a lattice gas model in $1$-d, we have obtained the steady
state distribution $P(m)$ exactly for small and large $w$. We have
further mapped the lattice gas model to an interface model in $(1+1)$
dimension and studied the width of the interface that characterizes
its fluctuations. We have calculated the roughness exponent $\chi$ and
the dynamical exponent $z$ analytically for small and large $w$ both
with and without bias. We have computed these exponents numerically at
the critical point and shown that the two critical points (with and
without bias) represent two new fixed points of interface dynamics in
$(1+1)$ dimension.

We have also generalized our model to include arbitrary fragmentation
kernels and obtained a few exact results for special choices of these
kernels via mappings to other solvable statistical mechanics models
studied in the recent past.

Our model obviously has some shortcomings as a model of realistic
aggregation and fragmentation phenomena. For example, we have assumed
that the rate of hopping of a mass as a whole is independent of the
mass.  In a realistic setting this is perhaps not a good
approximation. For realistic modelling of a specific system, one needs
to put in these details. This however was not attempted in this paper,
whose aim is to understand the mechanism of the dynamical phase
transition induced by the basic microscopic processes of diffusion,
aggregation and fragmentation within a simple setting.

However there remain many open questions even for this simple model.
For example, in one dimension we were not able to compute analytically
the various exponents at the critical point. Given that there has been
recent progress in determining the exact steady states of a class of
exclusion processes in one dimension via using matrix product
ansatz\cite{DE} and coordinate Bethe ansatz\cite{GS}, it may be
possible to obtain the exact steady states of our model in $1$-d.

Another important open question that needs to be studied for both the
symmetric and asymmetric models: What is the upper critical dimension
$d_c$ of these models beyond which the mean field exponents will be
exact. From general analogy to other diffusion limited models studied
earlier\cite{Cardy}, one expects that in the symmetric case
$d_c=2$. This expectation is supported by the numerical fact that the
the exponent $\tau_{s}\approx 2.33$ in $1$-d is already close to its
mean field value $5/2$. This conjecture however requires a proof.

The question of upper critical dimension is however puzzling for the
asymmetric model where the numerical value of $\tau_{as}\approx 2.05$
(perhaps $\tau_{s}=2$ with logarithmic corrections) is quite far from
the mean field value $5/2$.  This indicates that the $d_c$ (if there
exists one) for the asymmetric case might even be bigger than $2$
which however is quite contrary to the naive expectation that the
$d_c$ of directed models is usually lower than that of undirected
models. A numerical study in $2$ dimensions of the asymmetric model
may shed some light on this puzzle.
 
We thank D. Dhar and Rajesh R. for useful discussions.

\section*{Appendix:  Exact Result for Uniform Fragmentation Kernel}

In this appendix we solve exactly the steady state single site mass
distribution function $P(m)$ for the model where the mass $m_i$ at
every site $i$ evolves in discrete time according to the stochastic
equation,
\bea
m_i(t+1)= q_{i-1,i}m_{i-1}(t) + (1-q_{i,i+1})m_i(t)
\label{lange}
\eea
where the fractions $q_{i,j}$ are independent and identically distributed
random variables in $[0,1]$ with distribution function $\eta(q)$. We note
the formal similarity between Eq. \ref{lange} and the force balance 
equation in the $q$-model of Coppersmith et. al.\cite{SC},
\bea
W(i,D+1)=\sum_{j}q_{ji}W(j,D) +1
\label{Copper}
\eea 
where $W(i,D)$ represents the net stress supported by a glass bead at a
depth $D$ in a cylinder and $q_{ji}$ is the fraction of the stress
transported from particle $j$ at layer $D$ to a particle $i$ at layer
$(D+1)$. The only difference between Eq. \ref{lange} and Eq. \ref{Copper}
is in the additional constant term $1$ in Eq. \ref{Copper} that is absent 
in Eq. \ref{lange}. Nevertheless the same line of argument as in ref.
\cite{SC} leads to an exact solution in our case also as we outline below.

We first consider the mean field theory where we neglect correlations
between masses. For simplicity, we use the notations, $q_{i-1,i}=q_1$,
$1-q_{i,i+1}=q_2$, $m_{i-1}=m_1$ and $m_i=m_2$.  Mean field
approximation then leads to to the following recursive equation for
the mass distribution,
\bea
P(m,t+1)&=&\int_{0}^{1}\int_{0}^{1}dq_1dq_2{\eta (q_1)}{\eta (1-q_2)}
\int_{0}^{\infty}\int_0^{\infty}dm_1 dm_2 \nonumber \\
&\times& P(m_1,t)P(m_2,t)\delta (m-m_1q_1-m_2q_2), 
\label{evolve}
\eea
where we have used Eq. \ref{lange}. In the limit $t\to \infty$ and for the
uniform distribution $\eta(q)=1$, the
Laplace transform ${\tilde P}(s)$ of the distribution $P(m)$ satisfies
the equation
\bea
{\tilde P}(s)=\big [ \int_0^1 dq{\tilde P}(sq)\big ]^2.
\label{a1}
\eea
Defining $V(s)=\sqrt
{{\tilde P}(s)}$ and $u=qs$, we get from Eq. \ref{a1},
\bea
sV(s)=\int_0^s du V^2(u).
\label{a2}
\eea
Differentiation with respect to $s$ yields,
\bea
V(s)+s {{dV}\over {ds}}=V^2(s).
\label{a3}
\eea
which can be integrated to give,
\bea
V(s)={1\over {1-Cs}}.
\label{a4}
\eea 
The constant $C$ is determined from the mass conservation equation,
$\int_0^{\infty}mP(m)dm =-d{\tilde P}(s)/ds|_{s=0}=\rho$ where $\rho$ is
the conserved mass density. Thus, $C=dV/ds|_{s=0}=-{\rho}/2$. Hence we
get, ${\tilde P}(s)=1/(1+{\rho\over {2}}s)^2$ and by inverse Laplace
transform,
\bea
P(m)={4m\over {\rho^2}} e^{-2m/\rho}
\label{a5}
\eea

This is the mean field result for $P(m)$. However in ref.\cite{SC}, it
was proved that for uniform distribution $\eta(q)=1$, the mean field
stress distribution, where the stresses satisfy Eq. \ref{Copper}, is
exact. The explicit algebraic proof in ref.\cite{SC} proceeded via
constructing exact recursion relations for the joint probability
distribution of the weights in row $(D+1)$ in terms of those for row
$D$, and showing that the mean field factorization of these joint
distributions are invariant under this recursion. The same line of
proof can be adapted to show that the mean field result Eq. \ref{a5}
is also exact for our problem. We do not give details of the proof
here as they can be found in ref.\cite{SC}.

\end{multicols}
\newpage

\vspace{2cm}

\vbox{
\epsfxsize=8.0cm
\epsfysize=8.0cm
\epsffile{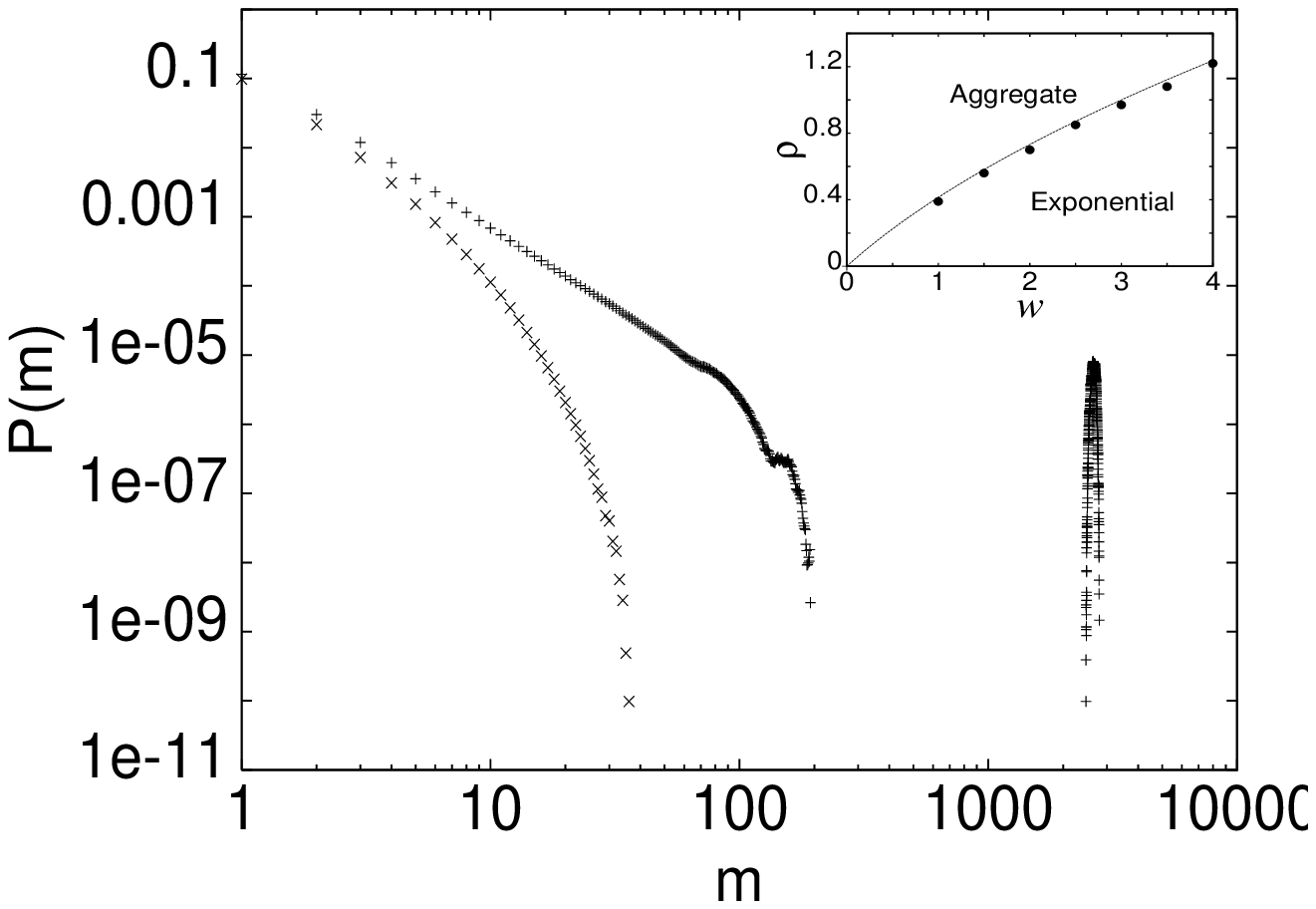}
\begin{figure}
\caption{ log-log plot of $P(m)$ vs. $m$ for SCMAM for two choices of the 
parameters: $(w=1.0, \rho=0.2)$ and $(w=1.0, \rho=3.0)$ on a periodic lattice
of size $N=1024$. Inset: Phase diagram. The solid line and the points 
indicate the phase boundary within mean field theory and
$1$-d simulation respectively.}  
\end{figure}}

\vspace{2cm}

\vbox{
\epsfxsize=8.0cm
\epsfysize=8.0cm
\epsffile{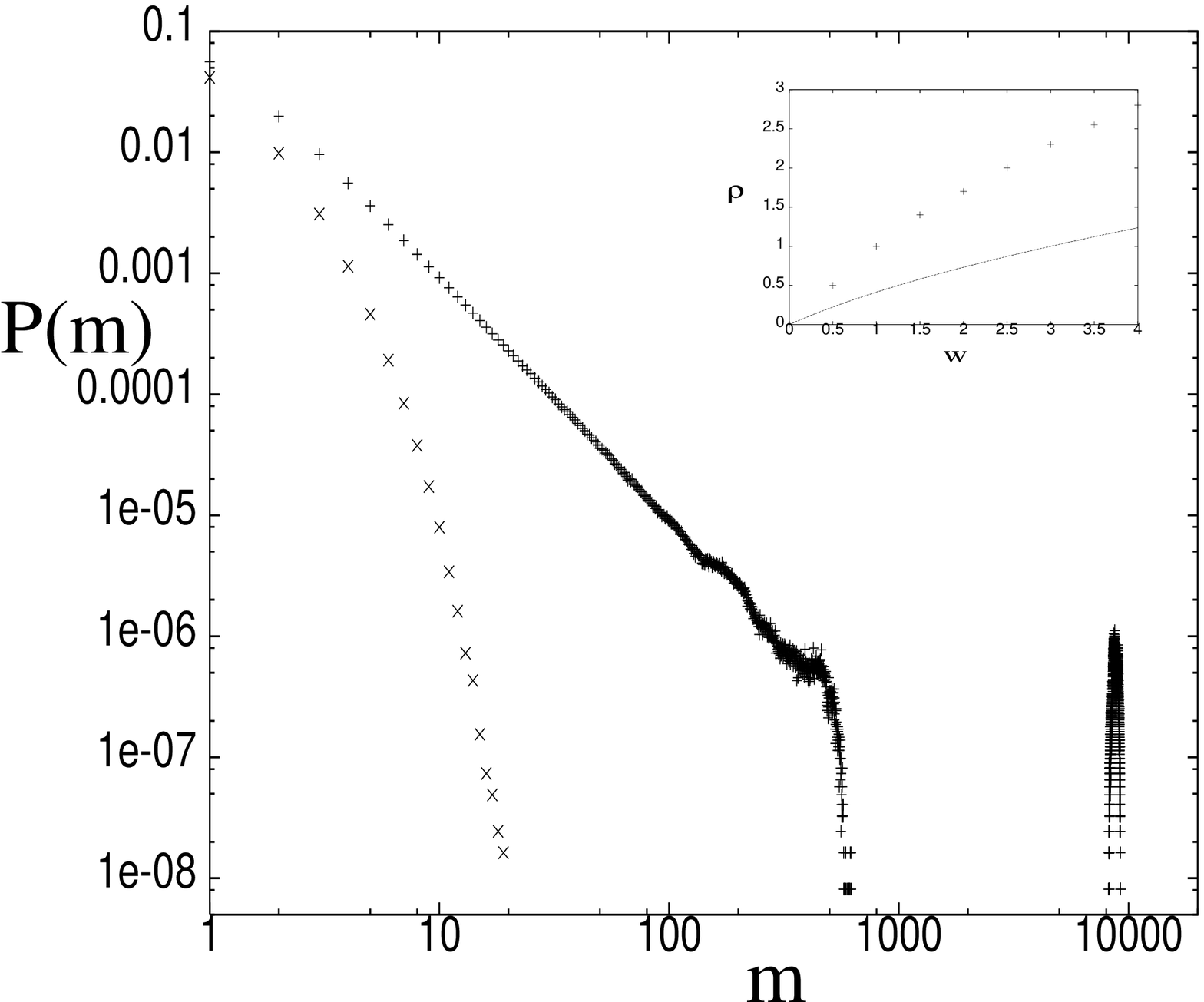}
\begin{figure}
\caption{log-log plot of $P(m)$ vs. $m$ for ASCMAM for two choices of the 
parameters: $(w=1.0, \rho=0.2)$ and $(w=1.0, \rho=10.0)$ on a periodic lattice
of size $N=1024$. Inset: Phase diagram. The solid line and the points 
indicate the phase boundary within mean field theory and
$1$-d simulation respectively.
}  
\end{figure}}

\vspace{2cm}

\vbox{
\epsfxsize=14.0cm
\epsfysize=14.0cm
\epsffile{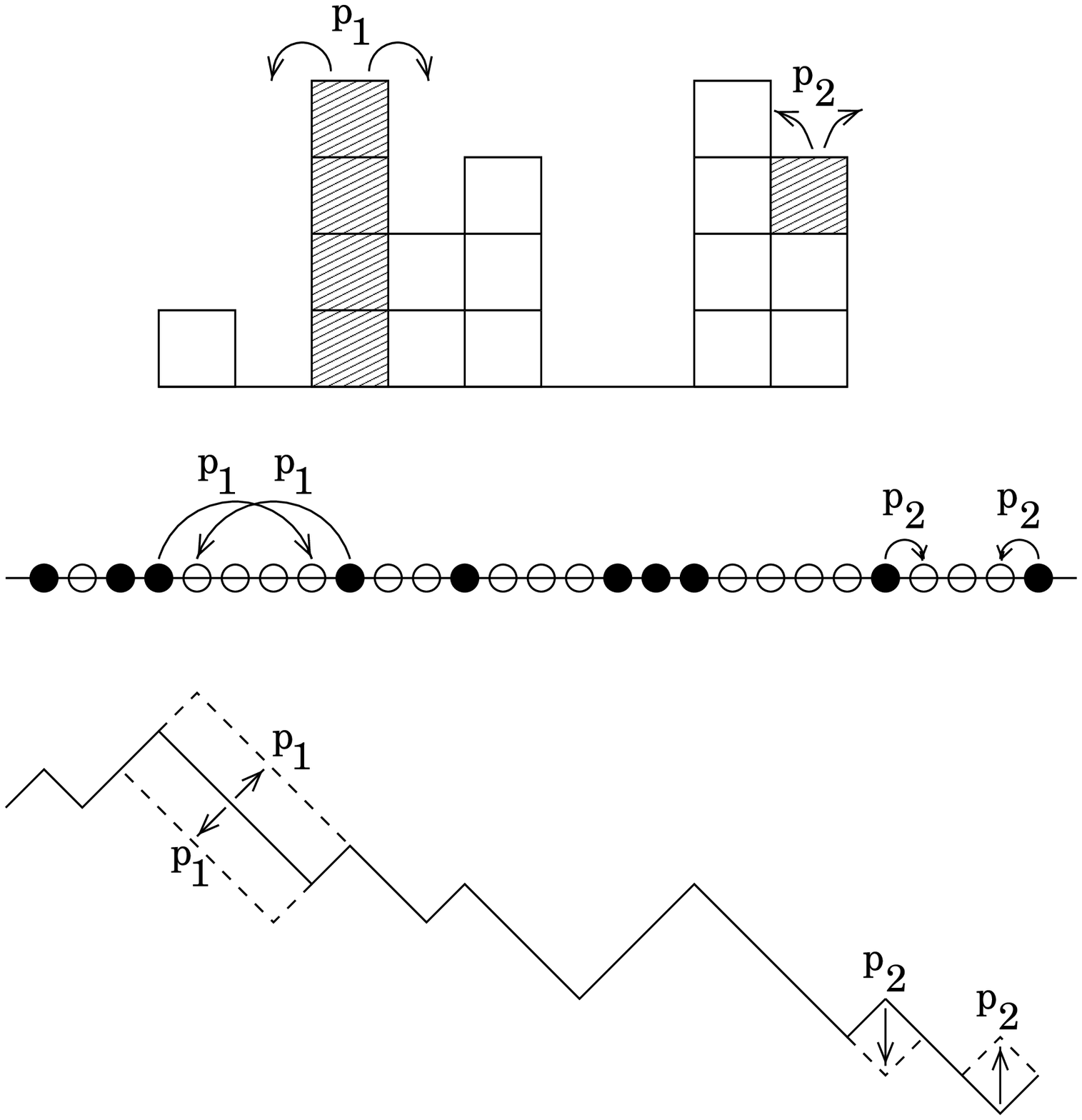}
\begin{figure}
\caption{The constructions of the equivalent lattice gas model and interface
model are illustrated for a particular configuration of the conserved mass
model. The shaded blocks indicate the masses that would move as a result
of ``diffusion and aggregation'' and ``chipping''.
}  
\end{figure}}

\vspace{2cm}

\vbox{
\epsfxsize=8.0cm
\epsfysize=8.0cm
\epsffile{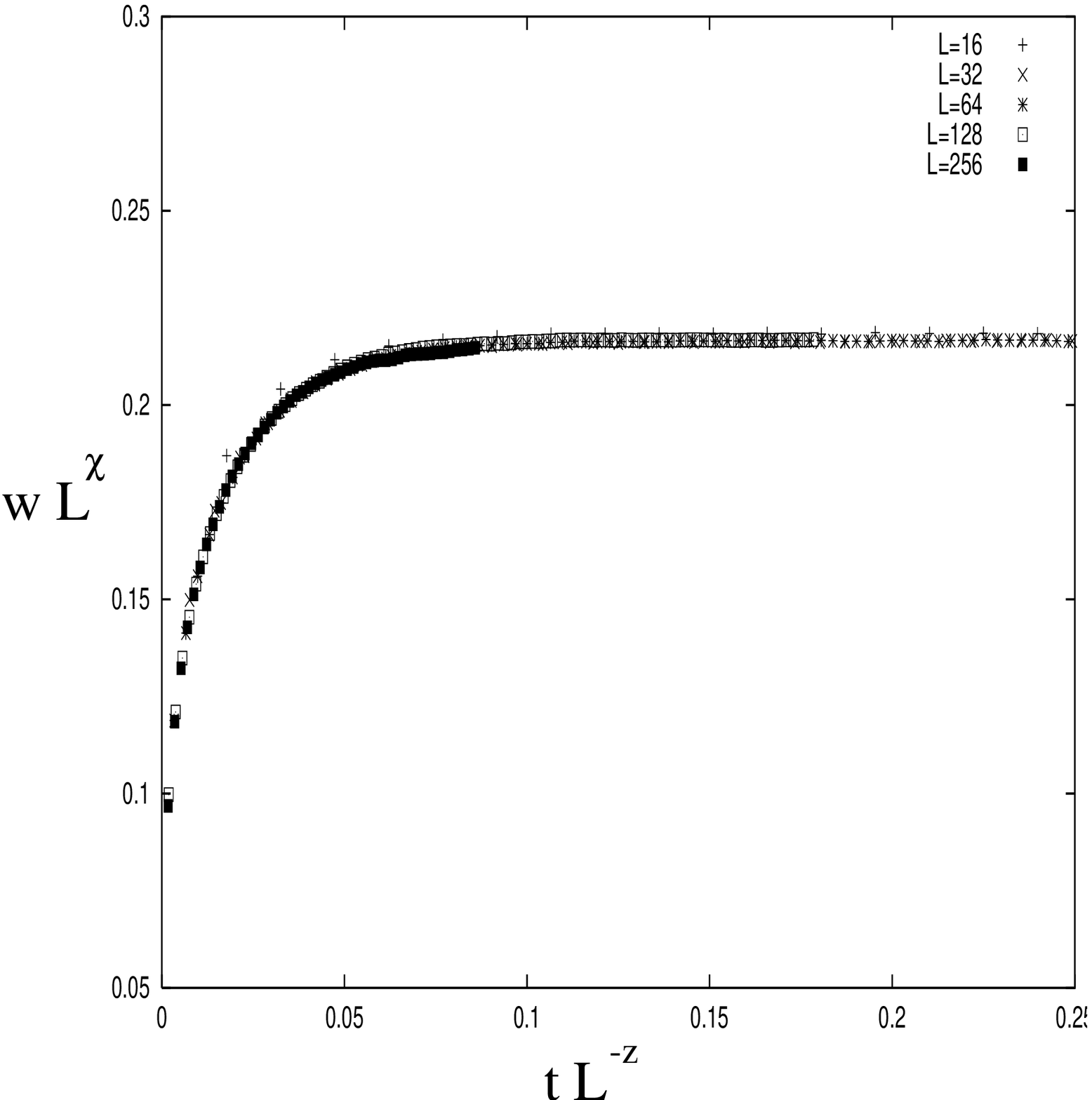}
\begin{figure}
\caption{ The scaled width $W/L^{\chi}$ is plotted against the scaled time 
$t/L^z$ for lattice sizes $L=16$, $32$, $64$, $128$ and $256$ at the critical 
point $(\rho=1,w_c\approx 3.35)$ of the interface corresponding to SCMAM. 
The best data collapse
is obtained with the choice of exponent values, $\chi\approx 0.67$ 
and $z\approx 2.1$.}  
\end{figure}}

\vspace{2cm}

\vbox{
\epsfxsize=8.0cm
\epsfysize=8.0cm
\epsffile{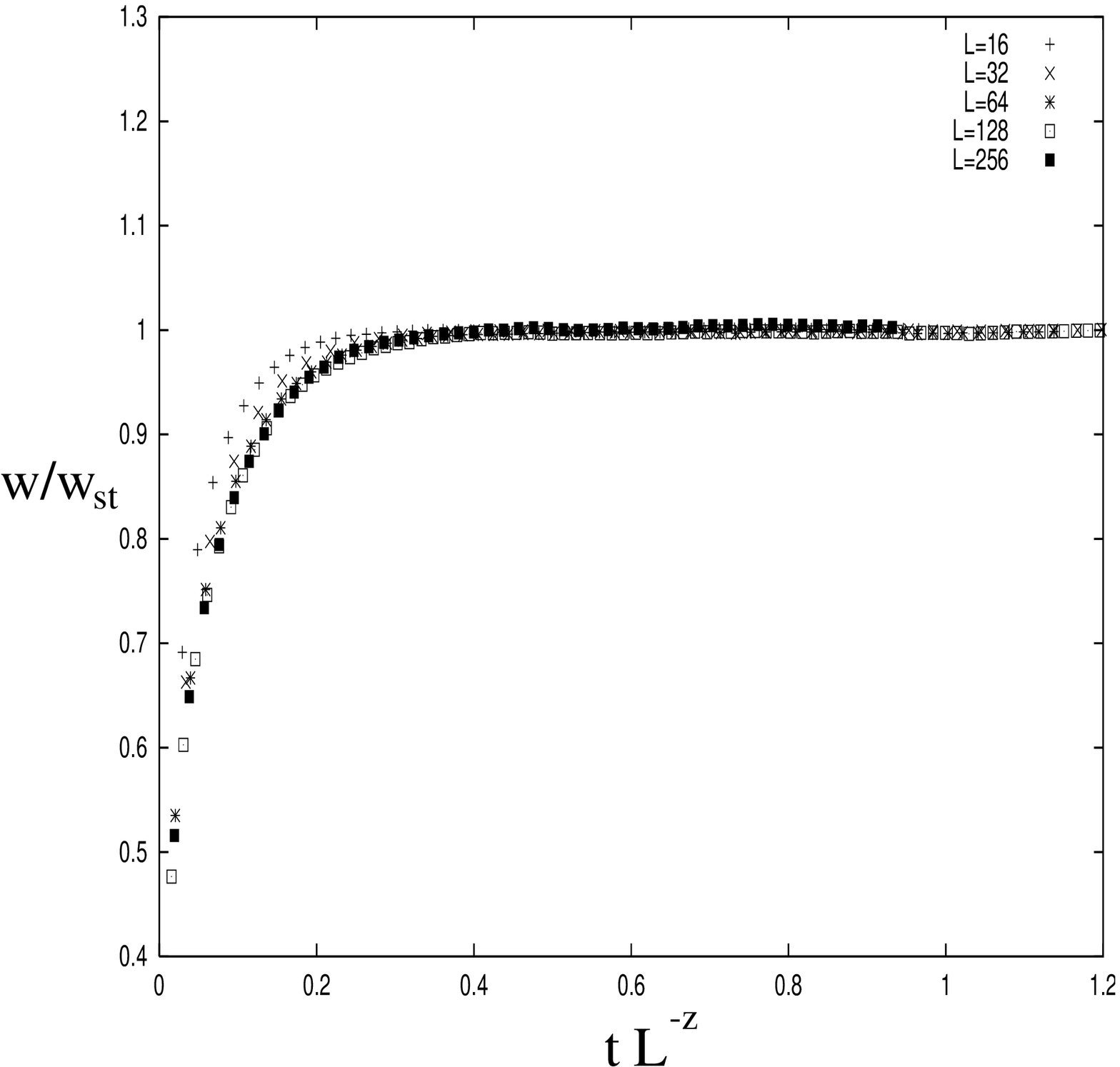}
\begin{figure}
\caption{ The scaled width $W/W_{st}$ is plotted against the scaled time 
$t/L^z$ for lattice sizes $L=16$, $32$, $64$, $128$ and $256$ at the critical 
point $(\rho=1,w_c\approx 0.77)$ of the interface corresponding to ASCMAM. 
The best convergence to data collapse as $L$ increases
is obtained with the choice $z\approx 1.67$. Inset shows $W_{st}$ plotted
against $L$.}  
\end{figure}}

\vspace{2cm}

\vbox{
\epsfxsize=8.0cm
\epsfysize=8.0cm
\epsffile{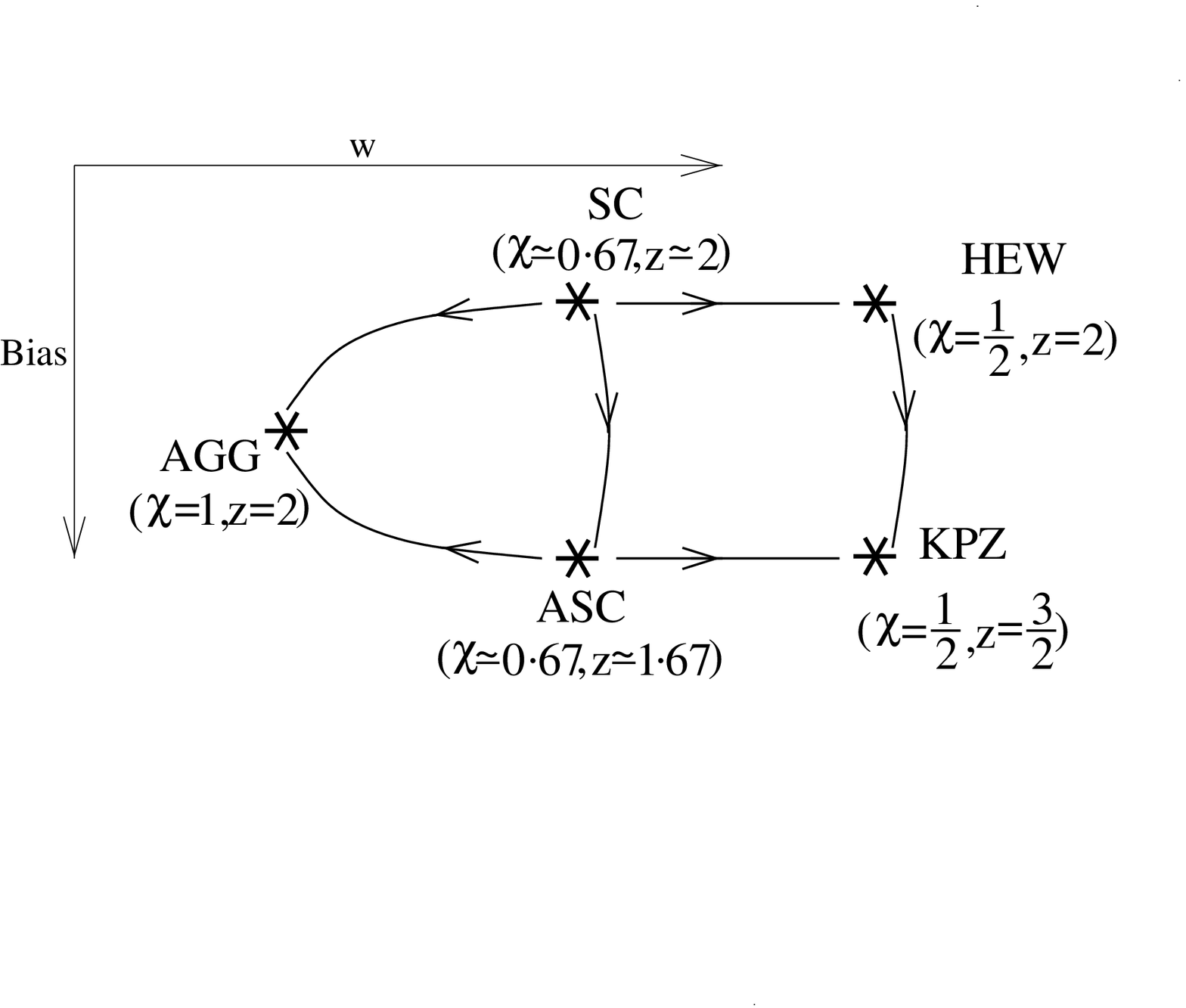}
\begin{figure}
\caption{Schematic depiction of fixed points and associated flows
}  
\end{figure}}

\vspace{2cm}

\vbox{
\epsfxsize=8.0cm
\epsfysize=8.0cm
\epsffile{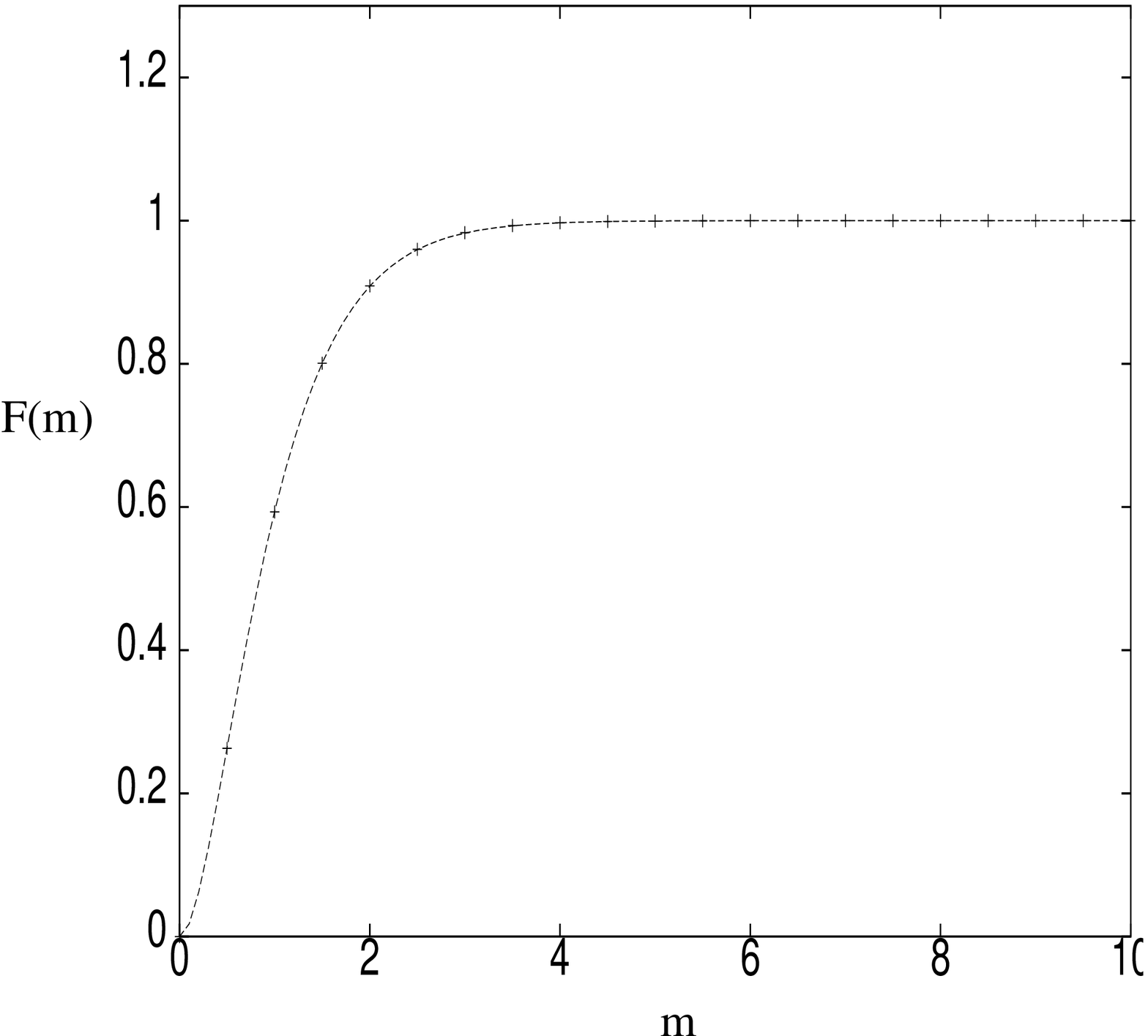}
\begin{figure}
\caption{ The theoretical cumulative mass distribution function,
$F(m)=1-e^{-2m}-2me^{-2m}$ (continuous line) plotted against $m$ along
with
the numerically obtained points from the simulation of the model
in one dimension for mass density $\rho=1$.}
\end{figure}}

\end{document}